# Molecular Size Dependence of the Field Effect in Organic Single Crystals

V.Y. Butko*, X. Chi, A. P. Ramirez
Los Alamos National Laboratory, Los Alamos, New Mexico, USA

We explore the effect of molecular size on injected carrier transport. We have fabricated and characterized field effect transistors (FETs) on optically transparent single crystals of tetracene, the molecule of which is comprised of four benzene rings (BR). These FETs exhibit effective channel mobility $\mu_{eff}$ up to 0.15 cm$^2$/Vs, intrinsic free carrier mobility of $\mu \sim 75$ cm$^2$/Vs, and on/off ratios up to $2\times10^7$. The present results show 1) the possibility of FET behavior in a transparent crystal with low intrinsic carrier density and 2) a weak dependence of performance on number of BRs, when compared to similar pentacene (5 BR) devices.

PACS: 72.80.Le, 71.20.Rv, 72.20.Jv

Semiconducting organic materials have received increased attention in the last several years because they promise bulk processing of flexible, large-area devices at low cost [1-4]. However, the molecular nature of organic systems also presents a different set of challenges when compared to inorganic systems. In particular, the energetic spectral bands of the molecular unit are extremely narrow, in addition to vibrational modes that have the potential to strongly scatter and even localize charge carriers [5]. This feature suggests a strong dependence of the electronic band structure, optical, and transport properties on the molecule type. On the other hand, though a wide variety of semiconductor organic materials have been studied, relatively little is known about the microscopic details of these effects on transport of injected into semiconductors charge carriers. Here we present a study comparing the behavior of Field Effect Transistors (FETs) fabricated from single crystals of tetracene and pentacene, molecules which have well-defined differences in molecular properties.

Among the simplest of organic systems are the polyacenes with general formula $C_{4n+2}H_{2n+4}$, where n is number of benzene rings in the molecule [5]. These molecules are generally flat and, due to their simplicity – the benzene rings form a linear chain – can be considered as model systems to study systematic effects due to molecular properties. The interest in the polyacenes also stems from the high value of room temperature effective mobility ($\mu_{eff}$) observed in polycrystalline organic thin film transistors (TFTs) made from thesemolecules. The effective mobility in pentacene films (n = 5), $\mu_{eff}$ = 0.3-3 cm$^2$/Vs [4, 6-8], is comparable to that of amorphous Si devices. However, pentacene, with an absorption region of 580 nm [5], is not transparent in bulk, restricting its use in optoelectronic applications. This focuses attention on the smaller polyacenes, n < 5, primarily because the band gap increases with a decrease in number of benzene rings. For instance, tetracene (naphthacene) (n=4) has an absorption region of 480 nm [5] and is transparent, making it a more suitable candidate for optoelectronic applications. Recently, fabrication and room temperature studies of polycrystalline tetracene TFTs with effective mobility, $\mu_{eff}$ = 0.1 cm$^2$/Vs were reported [9]. It is known that $\mu_{eff}$ increases with increasing crystallinity in pentacene TFT devices [3, 6, 7, 10], as expected from experience with inorganic semiconductors. In our previous work, we reported the fabrication and study of pentacene single crystal FETs [11]. In the present work we report on fabrication and characterization of FETs made from single crystal tetracene. Tetracene and pentacene both have the same triclinic crystal symmetry [12] and this similarity provides us with an excellent opportunity to study the effect of a molecular size on injected carrier transport.

The tetracene crystals used in the present study were grown by horizontal physical vapor transport in a stream of ultra high purity argon. The crystal growth apparatus was a modified version of the one reported by Laudise et. al. [13] with two glass tubes of different diameters. This is the same apparatus that we used to grow pentacene crystals [11]. The outer tube was wrapped with two rope-heaters that define the source zone and crystal growth zone, respectively. A glass tube of smaller diameter serves as the reactor tube. The source temperature was 245 C, and the flow rate was 19 ml /min. The source material was purchased from Aldrich and was once re-crystallized in argon for purification before use for crystal growth and also was crystallized without this purification stage. Typical crystal dimensions are 1-5 mm length, 0.2 – 5 mm width and 0.05 – 0.5 mm thickness.

In this work we use colloidal graphite for source and drain contacts. The graphite is painted on the smooth, untreated single crystal surface. The separation, L, between contacts is in range 50-150 μm and the width of the contact pads, Z, is 0.1-0.3 mm. Similar to the approach used in [11] on pentacene, the organic parylene was used as gate insulator (400-600 nm thick).

Parylene was deposited on the top of the crystal in a home-made reactor at a rate of less than 3 A/s. The thickness of the parylene layer is measured by Inficon XTM2 deposition monitor, calibrated from the capacitance measurements of planar devices fabricated simultaneously with the FET devices. Typical tetracene data represented in this paper have been obtained on the tetracene sample #10 with L = 50 μm, Z = 0.2 mm and parylene thickness, D = 650 nm. In the final step of FET fabrication, a silver paste gate contact electrode was painted on the top of the parylene over the region between the source and drain. The devices were not annealed. Current and voltage were measured with two Keithley 6517A electrometers. The electrical measurements were made in darkness in a Quantum Design cryostat at fixed temperature in a vacuum ~ $10^{-5}$ Torr. For these measurements, a 1-20 Volt step and a 5-30 second delay between each measurement was typical. Leakage gate current at the low voltages was ~$10^{-14}$-$10^{-13}$ A, and at the highest voltages applied, never exceeded $3 \times 10^{-12}$ A.

Typical transistor characteristics of devices at room temperature are shown in fig.1-3. The drain current ($I_{sd}$) exhibits a linear dependence on source-drain voltage ($V_{sd}$) at fixed gate voltages ($V_g$) (Fig.1) when $V_{sd} < V_g$. For $V_{sd} \geq V_g$ the current saturates at a near-constant value. This is standard behavior of a semiconductor FET [14]. In fig. 2 are shown $I_{sd}$-$V_g$ characteristics at fixed $V_{sd}$. One can see that the on-off ratio reaches $2 \times 10^7$ at $V_{sd} \sim$ -30 V. This on-off ratio is 4 times higher than we observed in single crystal pentacene FETs [11] and 2000 times higher than on-off ratio reported in rubrene single crystal FETs [15]. This higher on-off ratio in tetracene corresponds to a significantly lower current in the off state. From fig.3, which shows the dependence of $(-I_{sd})^{1/2}$ on $V_g$, we extract an effective mobility $\mu_{eff} \sim 0.15$ cm$^2$/(Vs) of the tetracene crystal at $V_{sd}$ = -50 V. This mobility value is by 50% higher than observed in a tetracene TFT [9], and two times less than pentacene single crystal mobility reported in [11].

Comparison of the observed mobility values of tetracene crystals with different purification procedures (this mobility difference was up to ~ 100 times) demonstrates stronger dependence of injected carrier transport on trap density than on the electronic band structure determined by size of the molecules. From fig.3 one can also extract threshold voltage values, $V_t$, of 15 V in tetracene and of 5 V in pentacene crystals. The threshold voltage corresponds to the gate voltage when the free carrier density equals the trap density [16]. Therefore, the higher observed threshold in tetracene leads us to a conclusion that the tetracene crystal used in the present study has a higher trap density than the crystals used in the pentacene study.

In Fig. 3 are shown results of a variable-temperature study of a tetracene single crystal FET at different values of $V_g$. We fit the data above the sensitivity limit of the electrometer ($3*10^{-14}$ A) to the form $R = R_0 \exp(-E_a/T)$, where $E_a$ is an activation energy, and also plot $E_a$ for the tetracene and pentacene crystals as a function of $V_g$ in fig. 5. We see that for slightly positive $V_g$, in room temperature range $E_{a0}$ ~ 0.56 eV and 0.67 eV for the pentacene and tetracene crystals, respectively. Both of these values are approximately 30% of the optical activation energy [5]. This is most likely due to an asymmetric distribution of traps which pin the Fermi level toward the valence band. For large negative $V_g$ (-50V), $E_a$ decreases to a significantly lower value, but still indicates thermally activated transport ($E_{a1}$ = 0.143 eV for the pentacene and 0.167 eV for the tetracene). This behavior appears to be different from both the TFT results which exclude thermally activated hopping as the fundamental transport mechanism in pentacene thin films [6] and the low temperature results of photoinduced carrier time-of flight (TOF) technique on naphthalene crystals, described by a band model [17]. However, it is very likely that carrier trapping or contact imperfection masks the intrinsic transport behavior in our crystals.

To reveal the fundamental free carrier crystal mobility, the analysis suggested in [11] can be applied. The observed behavior is well described in terms of a standard semiconductor model assuming that holes injected from the contacts become trapped at and below the Fermi energy level ($E_F$) inside the band gap [16, 18, 19]. In this picture $E_a \sim E_F-E_V$ ($E_V$ is valence band energy) and the dependence of $E_a$ on $V_g$ corresponds to a gradual shift of the hole Fermi energy toward the valence band as more empty shallow [20] [21] traps become filled due to FET hole injection. The number of thermally excited free carriers from these filled traps per unit area can be overestimated by the following [11]

$$p_f = \approx \exp(-E_{a1}/k_B T) C_s V_g. \qquad (1)$$

Here $E_{a1}$, $C_S$, $k_B$ and T are the activation energy with $V_g$ on, capacitance per unit area, Boltzmann constant and temperature, correspondently. From this equation and $E_{a1}$ = 0.167 eV at $V_g$ = -50 V, we find that at room temperature the number of free carriers in the tetracene crystal is about or less than 0.0015 of the total number of injected carriers. This fact along with $\mu_{eff} \sim$ 0.15 cm$^2$/Vs gives an estimation of a free carrier mobility in tetracene crystal $\mu \sim$ 100 cm$^2$/Vs. This value is the same order of magnitude as the one, $\mu_p \sim$ 75 cm$^2$/Vs, obtained in [11] for pentacene crystals. We note that $\mu_{eff}$ has been studied in the FET configuration in rubrene crystals [15] and $\mu$ by the TOF technique on other crystals [17, 22, 23]. The FET results on rubrene, $\mu_{eff} \sim$ 8 cm$^2$/Vs, are not inconsistent with our present results. The TOF results demonstrate a significant increase of intrinsic mobility above values of $\mu_{eff}$ reported for TFTs. However, a direct confrontation of FET and TOF results for the same material (and same batch) has yet to be made.

We report on successful fabrication of FET on the surface of tetracene single crystal. The obtained results such as high on-off ratio and high fee carrier mobility demonstrate that optically transparent tetracene is a suitable candidate for the optoelectronic applications. Comparing these results with those recently reported on pentacene crystals demonstrates stronger dependence of injected carrier transport on trap density than on the electronic band structure determined by size of the molecules at least at the present level of trap density. This observation emphasizes the importance of decreasing number of traps and improving the contact quality to reach a real free carrier injection into organics with $\mu \sim 100$ cm$^2$/Vs, which remains the ultimate goal of single-crystal device project.

We are grateful to D. Lang, V. Podzorov and G. Lawes for help and useful conversations and we acknowledge support from the Laboratory Directed Research and Development Program at Los Alamos National Laboratory.

* On leave from Ioffe Physical Technical Institute, Russian Academy of Science, Russia

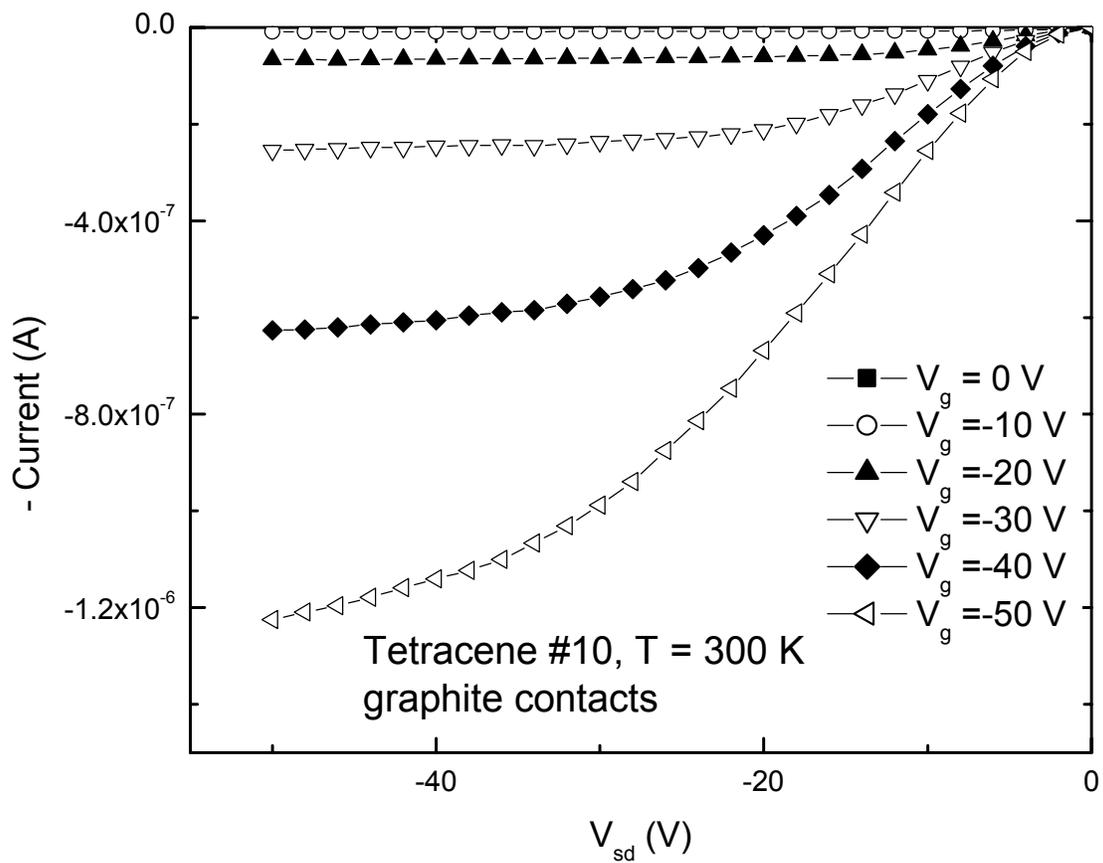

FIG.1.

$I_{sd}$ ($V_{sd}$) FET characteristics at different gate voltages.

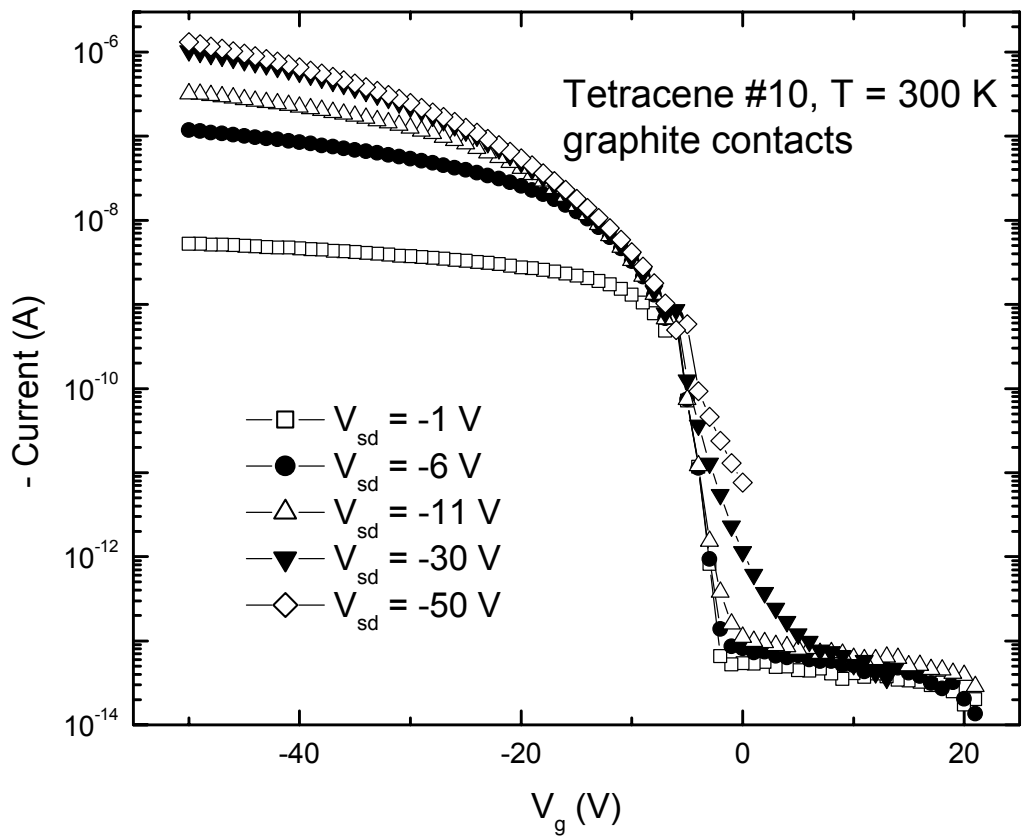

FIG.2.

-$I_{sd}$ ($V_g$) tetracene FET characteristics at different $V_{sd}$.

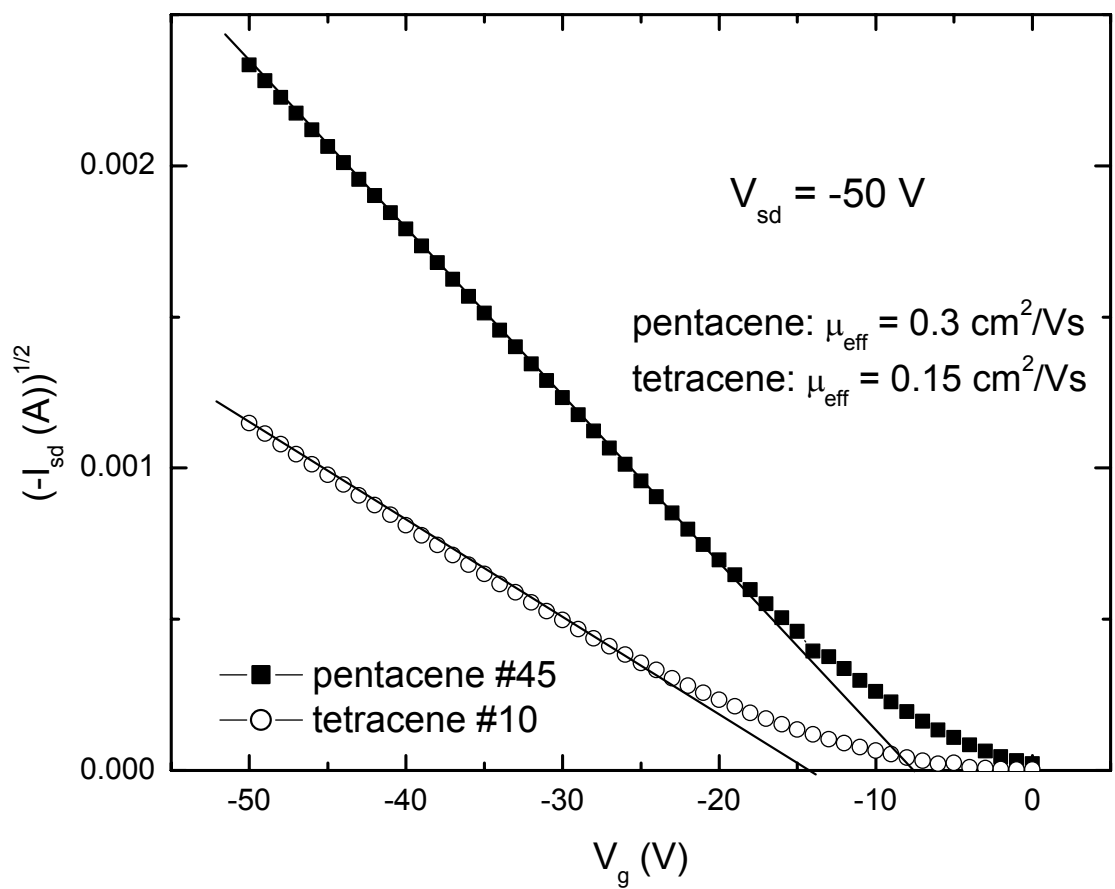

FIG. 3

$(-I_{sd})^{1/2}$ ($V_g$) pentacene and tetracene FET characteristics at $V_{sd}=-50V$.

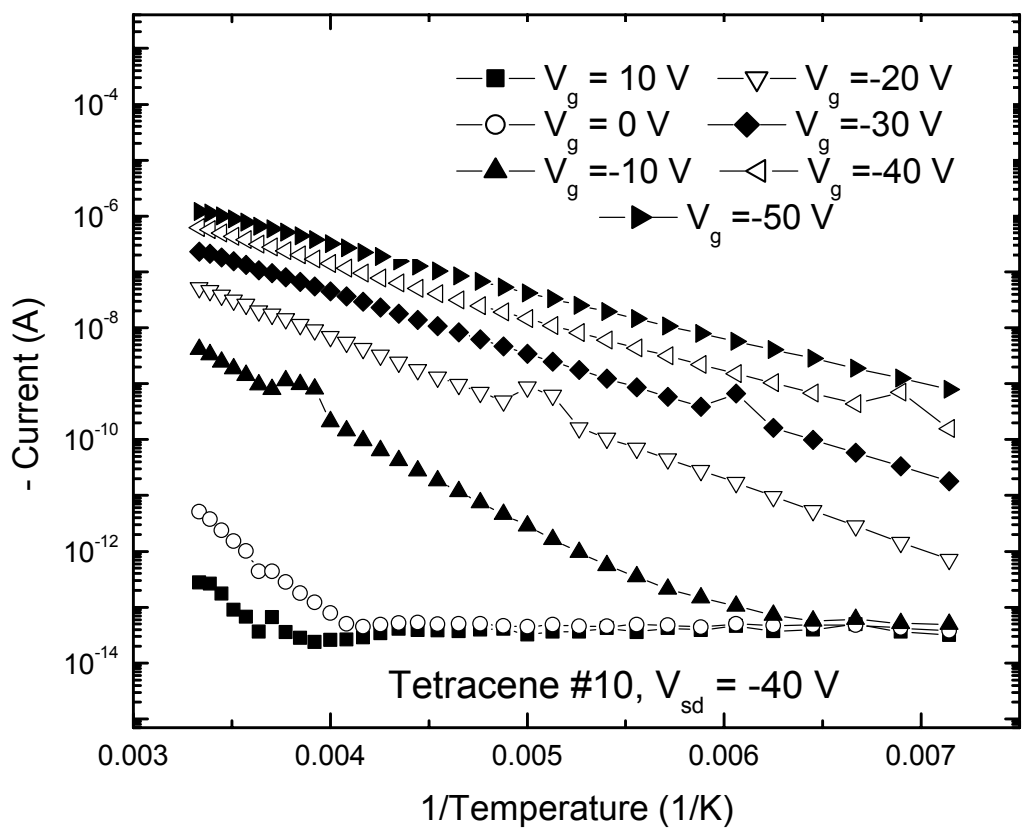

FIG.4.

Dependence of $-I_{sd}$ ($V_g$) on 1/Temperature at $V_{sd}$ = -40V in the tetracene crystal.

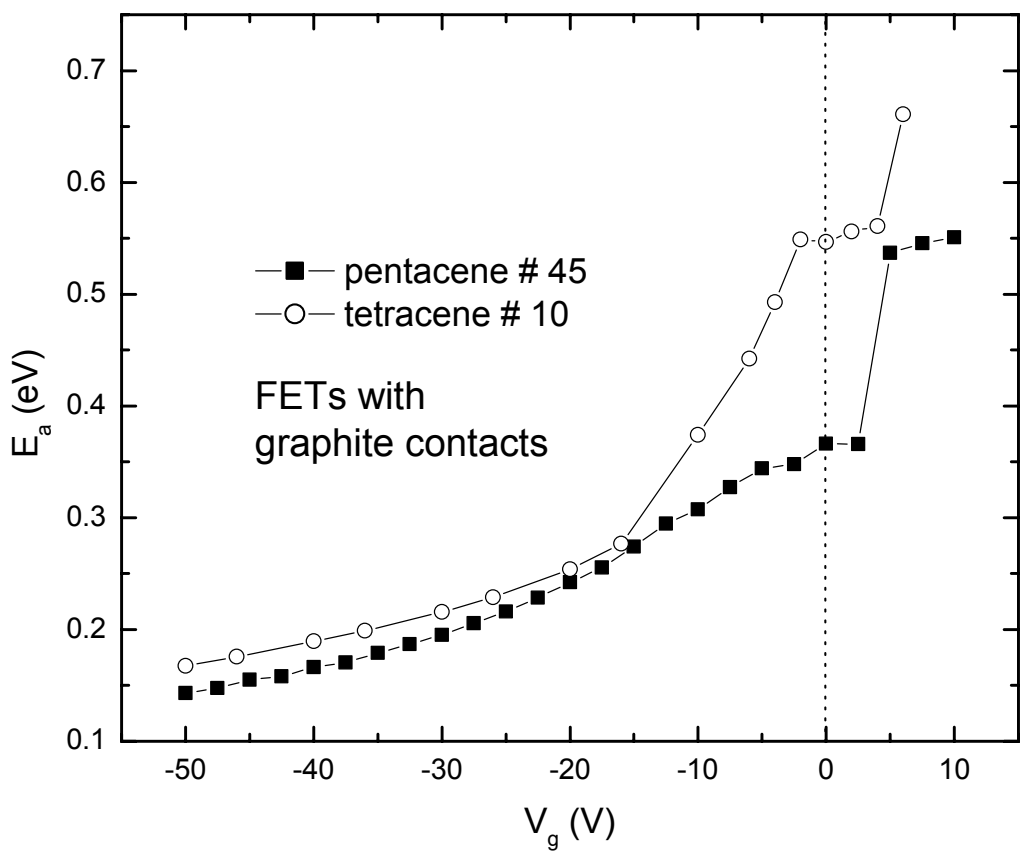

FIG.5.

Dependence of $E_a$ on $V_g$ in the tetracene and pentacene crystals in room temperature range.